\documentstyle[twoside,fleqn,espcrc2]{article}

\newcommand{\AmS}{{\protect\the\textfont2
  A\kern-.1667em\lower.5ex\hbox{M}\kern-.125emS}}

\title{Message passing on the QCDSP supercomputer}

\author{Michael Creutz \address{Physics Department, Brookhaven
National Laboratory, PO Box 5000, Upton, NY 11973-5000, USA\\
creutz@bnl.gov } 
\thanks{This manuscript has been authored under
contract number DE-AC02-98CH10886 with the U.S.~Department of Energy.
Accordingly, the U.S. Government retains a non-exclusive, royalty-free
license to publish or reproduce the published form of this
contribution, or allow others to do so, for U.S.~Government purposes.
The RIKEN/BNL/Columbia collaboration members include T. Blum, P. Chen,
N. Christ, M. Creutz, C. Dawson, G. Fleming, A. Kaehler, T. Klassen,
C. Malureanu, R. Mawhinney, S. Ohta, S. Sasaki, G. Siegert, C. Sui,
A. Soni, M. Wingate, P. Vranas, L.  Wu, and Yu. Zhestkov } }
 
\begin{document}

\begin{abstract}
The QCDSP machines were designed for lattice gauge calculations.  For
planning it is crucial to explore this architecture for other
computationally intensive tasks.  Here I describe an implementation of
a simple message passing scheme.  With the objective being simplicity,
I introduce a small number of generic functions for manipulating a
large data set spread over the machine.  I test the scheme on three
applications: a fast Fourier transform, arbitrary dimension SU(N) pure
lattice gauge theory, and the manipulation of Fermionic Fock states
through a distributed hash table.  These routines compile both on
QCDSP and a Unix workstation.
\end{abstract}

\maketitle

\section{Introduction}

The massively parallel QCDSP supercomputers located at Columbia
University and the RIKEN/BNL Research Center \cite{qcdsp} were
explicitly designed for large scale lattice gauge calculations.  The
machines are primarily run with software highly tuned tuned for four
dimensional lattices with an internal SU(3) gauge group.  As such,
they have effectively been serving as special purpose machines for a
single problem.

An open question is whether whether this architecture is sufficiently
flexible for more general tasks.  With this in mind, as well as with a
personal desire to explore the the machine, I developed a simple
message passing scheme.  My goal is a small number of generic
functions for manipulation of a large data set spread over the entire
machine.

For a machine to be ``general purpose'' has two prerequisites.  First
is a compiler in a higher level language.  This is provided by the
optimizing Tartan C/C++ compiler from Texas Instruments.  Second is an
efficient communication scheme between the individual nodes.  The rich
software environment of the Riken/BNL/Columbia collaboration provides
this for the primary application of the machine.  In this mode the
machine, while capable of MIMD operation, runs in a SIMD manner.  A
high degree of tuning obtains excellent performance, up to 30\% of the
theoretical peak speed of the machine.

In contrast, the aim of the project described here is a highly
flexible communication package for rapid prototyping of a variety of
problems.  In the process, some efficiency loss is expected.  I hide
the basic geometry of the machine from the top level, and applications
are developed entirely in a higher level language.  The source and
more details are available on the web \cite{mypage}.  The goal is
similar to but much less ambitious than the MPI project \cite{mpi}.

\section{Top level}

My top level interface is designed with simplicity as the primary
goal.  The usage begins with the definition of a basic data type, and
proceeds with a small number of routines for manipulation of a large
assembly of objects of this type.  For example, in the case of lattice
gauge theory the data type might be $SU(3)$ matrices.  After the basic
type is defined, the communication package is included, making
available several routines to manipulate such objects.  A call to the
function {\tt allocate(n)} sets up space for {\tt n} items of this
type.  The way the allocation is spread over the processors is meant
to be fully hidden.  In the case of lattice gauge theory one would
allocate space for the total number of links.

The scheme revolves about three basic functions to manipulate the
allocated objects.  First {\tt store(i, \& item)} stores the data {\tt
item} at the {\tt i}'th allocated location.  The complementary
function {\tt fetch(i, \& item)} recovers the item.  Any processor can
store or fetch any item, and need not know on which processor it is
stored.

After stacking up a number of stores or fetches, all processors call a
synchronizing function {\tt worksync()}.  This allows the
communication to proceed, with the data being passed until all
pending stores and fetches are completed.  While multiple
stores/fetches can occur simultaneously, there is no guarantee of the
order in which events are completed.  When {\tt worksync()} returns,
the machine is synchronized.  

For efficient loops over the variables, it is useful to know what data
is stored on the current node.  This is accomplished with the boolean
function {\tt onnode(i)}, which returns true if item {\tt i} is local.

In addition to the basic interface, there are several conveniences
available.  A variant of {\tt store()}, {\tt add(i, \& item)} adds the
new item to whatever is already stored in location {\tt i}.  This
improves efficiency in eliminating the need to fetch the old stored
value, which could be on a distant processor.  A variant of {\tt
fetch()} obtains multiple stored items in parallel.

A few utility functions are included, such as global sums and
broadcasts.  A function {\tt cmalloc()} attempts to malloc space in
the fast memory on the processor chip.  These and similar functions
will presumably eventually be built into the machine operating system.

To test these routines, I implemented a ``fast Fourier transform'', a
pure gauge code, and a Grassmann integration routine involving
manipulation of large Fock spaces via hash table techniques.  The FFT
code works by recursively subdividing the lattice, giving each half to
half the remaining processors.  Once a sub-lattice is assigned to only
one processor, the procedure is a standard FFT.  After assigning the
various tasks, the results are combined, which involves heavy
communication.  The dominance of communication makes the overall
process discouragingly slow compared to running on a workstation.
 
My pure gauge code is more satisfying, running at about 2/3 the speed
of the equivalent code of the RIKEN/BNL/Columbia collaboration.
However, it is extremely flexible, allowing an arbitrary number of
space time dimensions, each of arbitrary even size.  The group is an
arbitrary $SU(N)$.  Much of the communication speed is due to the
ability to fetch several neighbors at once using the multiple fetch
function.

The Grassmann integration implementation works particularly well.  The
algorithm is based on Ref.~\cite{grassmann}, and involves a large
distributed hash table, spread over all the processors.  Each
processor handles a portion of this table, sending stores non-locally
to randomly chosen other processors.  The efficiency is primarily due
to the parallel nature of the communication, and the fact that an item
in the process of being stored is not needed for immediate
computation.  The primary limitation of the algorithm is the
exponentially large amount of memory required, quickly exhausting the
limited amount on the current machine.

The distributed hash table uses simple extensions of the
communications class.  Instead of a single data type, two are used.
One, {\tt hindex}, is a type used to index the other, an {\tt hvalue}.
After defining these classes, including the file {\tt hashcom.C} in
turn includes the communication routines.  Storing an item uses {\tt
hstore(hindex,hvalue)}, while fetching involves the complementary {\tt
hvalue hfetch(hindex)}.  The function {\tt worksync()} is used as
before for the communication to proceed.  The storage is random over
the entire machine.

To manipulate the table, each processor handles his local part.  On
storing, the final location is unknown, but is not needed by the
algorithm.  Parallel loops over the table are fast since all
operations are carried out locally and the non-local storage proceeds
in parallel.  The processor needs only occasionally check for active
messages to keep the communication running.

\section{Middle level}
My goal was to keep the details of the communication as hidden from
the top level as possible.  The data is passed around in messages, the
basic message structure containing the identities of the source and
destination processors, one data element, a verb to indicate what to
do with the data (store, fetch, acknowledge, error, etc.), and an
extra word for various purposes, such as to carry the index of the
element.

The machine architecture is a four dimensional toroid with nearest
neighbor serial connections.  While these are in principle
bi-directional links, for simplicity I always send messages in one of
the positive directions. Each processor listens for incoming
messages on the negative wires.  Thus any particular serial
connection is used in only one direction.  The advantage is
simplicity, while the disadvantage is that the messages may not follow
the shortest path to their destination.  A store and acknowledge
combination between different processors circles the machine.

Given a message, a lookup table determines which wires lead closer to
the destination.  The first free one is used.  If none are free, the
message enters a queue.  In this scheme all wires can be
simultaneously active.  The route from one processor to another is not
predetermined, but progresses according to the currently available
wires.

At this level, several internal functions appear.  First, {\tt
sendmessage()} selects and activates a wire to start a message
traveling.  If no wire is available, the message is put in a FIFO
queue.  The complementary function {\tt readmessage()} checks the
incoming wires for a completed transmission and forwards messages not
for the current processor.  A function {\tt handlemessages()} calls
{\tt readmessage()}, performs any requested actions, sends
acknowledgements, and checks the message queue.  The function {\tt
worksync()} works by repeatedly calling {\tt handlemessages()} until
all unfinished stores and fetches are completed.

\section{Bottom level}

The basic communication works through the custom serial communication
unit (SCU) of the individual nodes.  Program initialization fixes the
SCU registers for the message size and sets the receive address
registers to buffers in ram.

To send a message, a write to a send address register starts the
transfer.  Monitoring progress uses a poll of the SCU status register.
This is all implemented in C/C++, without any assembly language.

The function {\tt worksync()} uses two (of three) global interrupt
lines available on the machine.  One flags unfinished stores/fetches.
When this line is set by all processors, a second interrupt
synchronizes the machine.  I also currently use the interrupt lines
for global ands, broadcasts and sums, but these will presumably
eventually be replaced by operating system functions.

\section{Summary}

I have described a simple interface to the QCDSP machines.  The goal
is rapid prototyping of new ideas in a high level language.  I
compromise efficiency for flexibility.  For most problems I expect a
loss of a factor of 2 to 3 in speed.  The test examples show varying
performance.  The FFT functions somewhat disappointingly; here all the
complexity is in non-local communication.  For a simple lattice gauge
algorithm the approach performs nicely, with more flexibility than in
the highly tuned production code.  Remarkably, for the Fock state
manipulations involved in evaluating Grassmann integrals, the
performance was excellent up to system sizes where inherent memory
limitations appear.

\end{document}